\documentclass[a4paper,fleqn]{cas-sc}
\usepackage[authoryear,longnamesfirst]{natbib}

\def\tsc#1{\csdef{#1}{\textsc{\lowercase{#1}}\xspace}}
\tsc{WGM}
\tsc{QE}
\tsc{EP}
\tsc{PMS}
\tsc{BEC}
\tsc{DE}

\begin{document}
\newcommand{\Nat}{\mathord{\textit{Nat}}}
\newcommand{\post}[1]{#1^{\bullet}}
\newcommand{\pre}[1]{{}^{\bullet}#1}
\newcommand{\tu}[1]{\stackrel{#1}{\to}}
\newcommand{\Mpp}{{\mathcal M}(P)}
\newcommand{\Mp}[1]{{\mathcal M}(#1)}
\newcommand{\Expr}{\mathord{\textit{Expr}}}
\newcommand{\Var}{\mathord{\textit{Var}}}
\newcommand{\Guard}{\mathord{\textit{Guard}}}
\newcommand{\In}{\mathord{\textit{In}}}
\newcommand{\Con}{\mathord{\textit{Con}}}
\newcommand{\Pnt}{\mathord{\textit{Pnt}}}
\newcommand{\Op}{\mathord{\textit{Op}}}

\newcommand{\Type}{\mathord{\textit{Type}}}
\newcommand{\true}{\mathord{\textit{true}}}
\newcommand{\false}{\mathord{\textit{false}}}
\newcommand{\skp}{\mathord{\textit{skip}}}

\newcommand{\ie}{i.~e. }

\let\WriteBookmarks\relax
\def\floatpagepagefraction{1}
\def\textpagefraction{.001}
\shorttitle{Modeling MOOC learnflow with Petri net extensions}
\shortauthors{IA Lomazova et~al.}

\title [mode = title]{Modeling MOOC Learnflow with Petri Net Extensions}                      
\tnotemark[1]

\tnotetext[1]{This work is supported by the Basic Research Program at the National Research University Higher School of Economics.}

\author[1]{Irina A. Lomazova}[orcid=0000-0002-9420-3751]
\cormark[1]
\ead{ilomazova@hse.ru}
\ead[url]{https://www.hse.ru/staff/ilomazova}

\author[1]{ Alexey A. Mitsyuk}[
auid=000,bioid=1, orcid=0000-0003-2352-3384]
\cormark[1]
\ead{amitsyuk@hse.ru}
\ead[url]{https://www.hse.ru/staff/amitsyuk}

\address[1]{HSE University, Moscow, Russia}

\author[1]{ Aliya M. Sharipova}[orcid=0000-0001-9011-3188]
\cormark[1]
\ead{amsharipova@hse.ru}
\ead[url]{https://www.hse.ru/staff/amsharipova}

\cortext[cor1]{Corresponding author}

\begin{abstract} 
Modern higher education takes advantage of MOOC technology. Modeling an education process of Massive open online courses (MOOCs) as a dynamic and multi-agent process is one of the challenging tasks. In this paper, Petri net extensions are investigated in the context of the learnflow modeling. It is shown how a learnflow can be modeled with  classical and colored Petri nets. These extensions facilitate modeling distributed and multi-agent processes. 
However, existing Petri net extensions do not provide the ability to model an education process in the context of multi-course programs and adaptive learning. 
We propose \emph{Petri nets with reference data} (PNRDs) for modeling e-learning in MOOCs. PNRDs allow us to represent a model of the education process in a visual, clear and not overloaded form. Moreover, PNRDs enable us to display aspects of multi-course programs and dynamic changes in the MOOC education process. We also show how PNRDs can be used to model online student collaboration in project-based learning.
\end{abstract}


\begin{keywords}
Massive open online courses \sep e-learning \sep  modeling multi-course learnflow \sep adaptive learning \sep Petri nets 
\end{keywords}

\maketitle

\section{Introduction}

Modern education takes advantage of information systems and data analysis \citep{AlRahmi2019}. The education process is supported with learning management systems (LMSs) which help educators to prepare teaching materials, organize the teaching process, and help students to organize their learning processes. MOOCs have become a well-adopted technology for self-education and additional learning. The growing number of MOOCs via various learning platforms is a testament to the fact that the role of MOOC technology in higher education and advanced studies is increasing \citep{Chen2020}. Currently, not only the increase in the number of courses can be noted, but also the number of participants using MOOC technology. Students pay attention to the university ranking, the efficiency, and usability of the online courses, it is therefore imperative to use flexible and modern teaching tools in MOOCs \citep{FidalgoBlanco2015}. 

The MOOCs are investigated from different perspectives: pedagogical, organizational, technological. The problems of content creation \citep{Hewawalpita2018}, the problems of evaluating work and confirming identity in MOOCs are considered from a pedagogical point of view. The following areas of MOOC research can be distinguished: the personalization of the learning trajectory, MOOC adaptability \citep{Pardos2017, Topali2019}, the prediction of  student performance, MOOC dropouts. Recent papers\citep{Garcia2018, SeinEchaluce2016} consider the necessity of the personalized learning and MOOC adaptability. The study by \citep{Qiu2016,Julia2021} researches the problem of students' motivation and engagement, \citep{ElMhouti2016} considers it in the context of reducing the dropout rate. The student grade prediction according to student forum activity is considered in \citep{Wise2018}. MOOC researchers utilize various approaches, such as data analysis \citep{Zhang2019, Jung2018}, psychometrics in an analysis of dropouts, and the prediction of student learning outcomes \citep{Fang2019, Reparaz2020, Abbakumov2018}. 
Also, there is an increase in interest in analysis of MOOCs processes not only from pedagogues, but also from software developers and managers. Thus \citep{Dumbach2020} discusses application of  process mining  for supporting e-learning information systems.

E-learning is a dynamic process by nature. It is supported by information systems. In this regard, the education process in MOOCs can be viewed as a classical business process, which is not different from manufacturing processes, sales, or any other dynamic multi-agent processes. The graphical representation of business processes or workflows is carried out by means of business process modeling to determine potential inefficiencies of a business process and improve it. The business process model is a description of a real business process, subsequently, data from the real education process can be analyzed. Results of such analysis can be helpful both for students and teachers. For example, the rules for studying the course or requirements for passing it. The business process model can be used for the analysis and development of LMS software, as well as the support of information systems, and is intended for monitoring real e-learning processes. It is crucial to consider not only the e-learning structure but also its dynamic properties for modeling online education processes. Related to education business process, MOOC models need to describe the control-flow of participants' interconnections recorded in LMS.  We focus on the presentation of a MOOC learning process as control-flow model taking into account its dynamics. 

The analysis of a student's learning trajectory can be based on workflow modeling. However, a demonstration of how the student moves from subject to subject in one course and the main result of the course study --- whether the student completed the program --- does not reflect the actual situation. In this case, the workflow model, where the student studies one course, is convenient to check the correctness of a workflow.  Nevertheless, many essential aspects are not taken into consideration. In reality, a student may take more than one course at once. It is possible to construct a model for the case of ``one student--one course'' with an approved set of prerequisites. A student plans to study in course ``A''. However, to enroll, the student needs to have the skills from course ``B''. While studying the course, the student could have missed assignments, study some topics, thus moving forward in his/her learning path. This means \emph{adaptive learning}, because we check not only compliance with the prerequisites for enrolling in the course, but also take into account the changes in the student's state in the learning path that have occurred in the dynamics. These changes are not taken into account in the model employing basic workflow modeling, however. In this case, the MOOC is not a single workflow, but a system of interacting workflows. Moreover, by using the model with references to data, the courses a student has taken can be identified and the records in the student's portfolio are updated. 

Another case from the perspective of adaptive learning is where the student studies a common subject in parallel in several studied courses. A modern university is moving to an education format that assumes the possibility of a flexible program transformation according to study interests. Course subjects do not necessarily correspond to a standard program structure --- a set of prerequisites  \citep{Yu2017}. However, in online education, various programs may have the same subjects and a student may have the competencies and final grades of a completed program. This shows the need to model the education process, in the context of the adaptability of the student's learning trajectory, to his or her requirements. The construction of the program trajectory, with respect to individual student characteristics, can be reflected in his or her portfolio as corresponding records of completed programs and subjects, competencies, etc.

An important characteristic of the education process is its multi-perspective nature. For example, everyday classical or online university activities can be seen from the following viewpoints: student (undergraduate), professor/instructor, teaching assistant, education program administrator, graduate (PhD), student, research staff, etc. Persons with these roles are involved in the education process and have their personal concerns. Thus, each viewpoint has specific characteristics. Separate aspects of the education process can be modeled and analyzed based on separate views. For example, it is possible to have an independent model of course prerequisites, which restricts the number of courses available for a student and their possible order. There are aspects of education that cannot be modeled using only one view. 

Nowadays, project activities play an important role, especially in the education of specialists of ICT and software development \citep{Konak2019}. It is crucial to involve students in the entire development cycle (project) for their ability to see the whole process from various perspectives.
A project-based education is common for modern software and IT curricula. Students perform projects (usually) in small groups. 
This approach brings academia and industry closer to each other. The interaction between students in a group, their interaction with the instructors, and inter-group communications are all important to correctly model this project-based education process \citep{Papanikolaou2010}. 
 
In this work, we focus on developing a formalism to modeling the education process environment as a dynamic multi-agent system. Accordingly, it is possible to set the following tasks requiring consideration: 

\begin{itemize} \item Identifying aspects of MOOC modeling that were not previously addressed in the literature;
	\item  Modeling the learning process in MOOCs as a standard workflow;
	\item Using the extension of Petri nets to model the process taking into account adaptive control (dynamics);
	\item Modeling of teamwork on a project during MOOC study.
\end{itemize}

One of the well-known approaches of modeling and analyzing dynamic process are Petri nets. Petri nets have precise mathematical semantics and, hence, the correctness of a modeled process can be checked. Petri nets also have graphical notation, which facilitates their visualization and analysis. Note also that Petri nets underlie many applied languages for modeling business processes.

Classical Petri nets are applied to simple system modeling. Currently, there are a large number of different Petri net extensions, which, in one way or another, expand the functionality of the model. Extensions of Petri nets have been developed in response to the increasing demand for different model perspectives of distributed and multi-agent systems. The following extensions of Petri nets are known: Petri nets with time, colored Petri nets (CPNs), reference nets, nets-within-nets. In this paper, we show how colors and data references can be useful to model the considered aspects of the e-learning process. 

In CPNs,  tokens carry data values of different domains. In classical Petri nets, all tokens, by definition, are of the same type and, therefore, indistinguishable. However, in systems that we want to model, tokens are often different objects. In a CPN, each token has its own meaning (color) which makes it possible to distinguish tokens \citep{JensenBook}. The token value can be of a basic or any complex type. 
Another Petri net extension is reference nets. Reference nets are high-level Petri nets incorporating the concept of nets-within-nets along with Java inscriptions \citep{Kohler2001}. Java inscriptions are used to create nets with bindings to reference variables. These references can be used to carry out communication between net instances. The transition firing can also be controlled through ``guard'' conditions \citep{Kummer2004}. 

For various tasks, Petri net extensions are used to capture all aspects of the concurrent distributed processes \citep{Carrasquel2019}. In our case, it is important to use the possibilities of building a multi-agent model, since the online education is a distributed process with many changes in real process. In reality, the model may be inconvenient for analysis, due to a large number of elements. In such a way, we propose utilizing the idea of reference tokens, where tokens are reference objects \citep{Carrasquel2019}, for developing multi-agent models of an e-learning environment. In this paper, we present Petri nets with reference data token (PNRD) --- an extension of the Petri net model with tokens allowed to be references which leads to very clear and intrinsic semantics.

Various problems of MOOC modeling are investigated to solve problems in different e-learning domains. However, this research does not answer the question, how to design the education process concerning adaptive learning. Our research aims to propose a formalism for modeling e-learning processes. This new formalism has to capture the multi-perspective nature of e-learning processes in MOOCs where students are free independent agents. This formalism enables us to model the e-learning process in the context of dynamic adaptability. This new extension can be used for modeling the multitasking of student learning, when the student may study many courses in parallel. Also, the Petri net extension provides the approach for modeling MOOC with teamwork.

The paper is organized as follows. Section~\ref{relatedwork} provides a review of related research on MOOCs and modeling education processes with a focus on Petri net semantics. In section~\ref{mainpart}, examples of Petri net modeling of MOOC study are presented. First, a model of studying in a MOOC by one student as a classical workflow model is provided. Then, we present how the extensions of Petri net can be utilized for solving the problem of the individual presentation of each student's data. If the student is studying one course, and we want to take into account the data, then it is possible to use CPNs. Examples of Petri net and their extensions and models for studying in MOOCs are presented as well. The third example is a model concerning dynamic adaptability, where the reference semantics extension was utilized. An example of the  teamwork workflow model is presented.  Section~\ref{discussion} analyzes the implications of the proposed Petri net formalism for modeling the education process and concludes the paper. 

\section{Related Work: Workflow, Learnflow and Multi-Agent Systems}\label{relatedwork}

The modeling and analysis of MOOC processes can become the basis for determining the main directions of course development, their structure, connections between courses, and the creation of a synergistic effect, and obtaining new knowledge for the development of education technologies \citep{Guerrero2020}. The e-learning environment modeling is a way to understand how various MOOCs aspects are related and what can be improved in order to increase the effectiveness of courses in terms of both knowledge and analytical results.
The  workflow modeling approach  \citep{Aalst2004} has been adopted for the learning process. Not long ago, the term ``learnflow'' \citep{Bergenthum2012} was proposed for the workflow of learning process. An actor learnflow model is a type of high-level Petri net where actors represent students, and roles are associated with transitions which represent process actions. These nets can be used to model a single group of students with different roles/competencies performing a project together. 

The enhancement of e-learning technologies has driven the education environment to meet requirements to improve and make more efficient and convenient the implementation of the MOOC technology \citep{Deng2019}. The increasing number of publications on LMS and MOOC consider various aspects of e-learning. The high level of dropout rates is a large part of the research in the MOOC analysis. The motivation of some learners in MOOCs is not always to complete the course \citep{Deeva2018}. The problem of MOOCs dropouts is considered in  \citep{Mubarak2020, Prakash2018} to explain what kind of problems the student can face  studying in a MOOC, which leads to a break in studies or a complete dropout. The curricular analytics from learning platforms and LMSs are used to explain the reasons for the high late dropout rate \citep{Salazar2021}.

The specific focus of the paper is on modeling education processes in MOOCs with the utilization of Petri nets. In the literature, related to an analysis of MOOCs study with Petri nets and its extensions utilization, the following reserch areas can be distinguished.

First, research focuses on the problem of the individual learning trajectory and adaptive learning environment. In this regard, the development of appropriate learning content according to students' preferences and requirements is analyzed \citep {Premlatha2015}. The development of an adaptive learning environment enables the construction of an individual trajectory for improving student's performance. Personalization of the student's learning is possible through learner classification to identify changing learner characteristics and requirements \citep{Premlatha2016}. 

In \citep{Vidal2013}, the problem of the course content adaptability according to the learner's performance is considered in the context of unified workflow patterns using an ontology-based and Petri net-based engine. The adaptability is considered in the context of making subsequent changes to the course structure. The authors of \citep{Hammami2015} consider the probability of adapting to the learner's preferences in the e-learning environment and analyzing and controlling the communication and interactions among the different agents in the context of the multi-agent nature of e-learning. The formal model of message exchanges on the basis of object Petri net is proposed.

Secondly, student MOOC performance is considered for the further prediction of the students' learning behavior and for blended learning. Student learning outcomes on the basis of students' activities in the e-learning system could be implemented as a valid indicator of student performance and its prediction for making a more efficient course  \citep{Conijn2018}. \citep{Balogh2019} discuss a simulation tool using Petri nets to predict student grades based on the use of learning materials.

Furthermore, the improvement of MOOCs is another problem that researchers focus on. Student performance analysis often provides researchers with the idea of evaluating the usefulness of course content and how to make improvements in MOOCs. A Petri net model can be applied for course development in respect of prerequisite connections \citep{Juhasova2016}. Hence, the fact that during the education process a student may do several courses at the same time is not considered. Balogh et al. \citep{Balogh2014} discuss the implementation of Petri nets as an effective tool for modeling the education process, and the model verification using various information analysis methods, such as usage analysis on the basis of log files. The event log data are used to understand which materials are being used by or are useful for students, and which can be excluded from the course in order to improve the student's learning within the course. 

Finally, it can be concluded on the basis of the literature review that the majority of investigations of e-learning modeling by means of Petri nets study mostly an individual learning trajectory or the prediction of student behavior. Current research usually focuses on individual student activity in MOOCs while disregarding the teamwork of students. There is a lack of papers focused on collaborative learnflow modeling. \citep{FroschWilke2008} have drawn attention to the fact that dynamic changes during collaborative learning processes should be considered on the basis of the reflective framework. However, despite the possibility of making changes in the model without redeveloping it entirely according to the changed factors, the availability of making changes related to students' teamwork on projects is not discussed, nor is the problem of process adaptation to changed external factors. Conversely, the problem of the adaptability of e-learning according to learner's level is not considered given that several courses may be taken at the same time. It is crucial for the model to reflect aspects of the interaction of prerequisites of a particular course and the possibilities of model development to include an adaptive student learning trajectory. 

The problem of adaptive learning in MOOCs is considered predominantly from the individual's characteristics such as a personality type, background, expectations. However, adaptability in the sense of dynamic changes is not taken into account. Although many articles model the education process in MOOCs, the models of e-learning as a distributed environment are scarce. 

The possibilities of process mining techniques and a large amount of LMS event log data enable the design of a model of the e-learning process in MOOCs, taking into consideration the system complexity and the distributed dynamic e-learning environment. 

In works devoted to multiple instances of process modeling and data analysis in Petri net, a distributed process is considered not an isolated process, but as data about objects that change over time. \citep{Fahland19} consider the modeling dynamics over multiple processes to avoid data extensions. In the paper of Ghilardi et al. \citep{Ghilardi2020} an extension of colored Petri nets, called catalog and object-aware nets (COA-nets) is utilized for multi-case and data-aware process modeling and analysis. An analogy can be drawn with the education process in MOOCs, where a student is in constant interaction with other participants in the process, that is, there are constant changes, the process is not discrete, moreover, it is necessary to take into account the changing student portfolio.

Modern MOOCs combine innovative project-based approaches with teaching elements of the traditional education format. For example, they propose different types of group or team projects. Teamwork is one of the most important competencies for most professionals, and it is therefore imperative that it be included in university programs \citep{Planas2020}. The integration of such courses into school/university academic programs is considered an important task to improve the quality and accessibility of education. It is also vital to model students' teamwork in MOOCs to analyze organizational bottlenecks and predict potential problems in order to improve online courses. 

In our opinion, it is essential to include in the education environment model all aspects of the education agents' interaction. These interconnections are recorded in LMSs as event logs. Therefore, it is important to indicate all the student's individual portfolio data on completed courses and final grades for more efficient student learning trajectory formation and avoiding the redundancy of studying previously completed courses.

In order to model multi-agent e-learning trajectories of different students, we considered reference semantics in Petri nets. It has been shown that reference nets suit the modeling of general social multi-agent systems \citep{Kohler2005}. The strength of reference nets is that this model supports a multi-perspective view of the modeled process. In this paper, we investigate how modeling using PNRD can be applied to the education process.

\section{Modeling the Education Process with Petri nets}\label{mainpart}

It is crucial to consider not only the e-learning structure, but also the features of its dynamic properties for modeling online education processes. In this paper, we focus on developing a formalism to support the design of the education process environment as a dynamic multi-agent system. Petri nets are intended for the visually modeling of distributed dynamic processes. Using mathematical tools Petri nets provide the possibility of model execution and verification.

\subsection{Modeling Subject Learning with Workflow Nets}
The process of studying a subject can be viewed as a business process in which  a student is a case that should be carried from admission to successful completion or drop out. 
To model business processes, a special class of classical Petri nets, called \emph{workflow nets} (WF-nets), is used.

We start with defining Petri nets and WF-nets. Then we will give a simple example of modeling the process of a subject study using a WF-net. 

A Petri net model is represented as a bipartite graph with nodes of two types:  places and transitions, which correspond, respectively, to local states/conditions/resources and actions/tasks/events of the system. Places may carry tokens, representing resources, control threads, etc. Thus, a current state of the system is represented by some distribution of tokens, called a Petri net marking. The dynamics of the system is modeled by  firings of transitions --  a transition firing changes the current state for a new one, i.e. it may change the number and  distribution of tokens.

Now we define Petri nets formally. 

Let $S$ be a set. A \emph{multiset} $m$ over a set $S$ is a mapping: $m\ :S\to \Nat$, where $\Nat$ -- is the set of natural numbers (including zero), i.e. a multiset may contain several copies of the same element. 

For two multisets $m,m'$ we write $m\subseteq m'$  iff
$\forall s\in S: m(s) \leq m'(s)$ (the inclusion relation). The
sum, the union and the subtraction of two multisets $m$ and $m'$ are defined as usual:
$\forall s\in S: (m+m')(s)=m(s)+m'(s), (m\cup m')(s)=max(m(s),m'(s)), (m-m')(s)=m(s)\ominus m'(s)$ (where $\ominus$ denotes the truncated subtraction).
By ${\cal M}(S)$ we denote the set of all finite multisets over $S$.

Let $P$ and $T$ be two disjoint finite sets of places and transitions respectively and $F:(P\times T)\cup (T\times P)\to \Nat$. Then $N=(P,T,F)$ is a \emph{Petri net}. 
A \emph{marking} in a
Petri net is a function $m: P\to \Nat$, mapping each place to some
natural number (possibly zero). Thus a marking may be considered
as a multiset over the set of places. A \emph{marked net} $(N,m_0)$ 
is a Petri net $N=(P,T,F)$ together with some initial marking $m_0\in\Mp{P}$. 

Pictorially, $P$-elements
are represented by circles, $T$-elements by boxes, and the flow
relation $F$ by directed arcs. Places may carry tokens represented
by filled circles (black dots). A current marking $m$ is designated by putting
$m(p)$ tokens into each place $p\in P$. Tokens residing in a place
are often interpreted as resources of some type consumed or
produced by a transition firing.

For a transition $t\in T$ an arc $(x,t)$ is called an \emph{input
	arc}, and an arc $(t,x)$
--- an \emph{output arc}; the \emph{preset} $\pre{t}$ and the
\emph{postset} $\post{t}$ are defined as the multisets over $P$
such that $\pre{t}(p)= F(p,t)$ and $\post{t}(p)= F(t,p)$ for each
$p\in P$. A transition $t\in T$ is \emph{enabled} in a marking $m$
iff $\forall p\in P\ m(p)\geq F(p,t)$, \ie   in its input places, $t$  has enough resources for firing. 
An enabled transition $t$
may \emph{fire} yielding  a new marking $m'=_{\mbox{\small def}}
m-\pre{t}+\post{t}$, i.~e. $m'(p)= m(p)-F(p,t)+F(t,p)$ for each
$p\in P$ (denoted $m\tu{t}m'$, or just $m\to m'$). 

A Petri net run is a sequence of firings $m_0\to m_1\to \dots$ starting from the initial marking $m_0$. Note that multiple transitions can be enabled at the same time in a marking $m$. Then the transition to fire is chosen non-deterministically.  A run can be infinite, or finite and end with a marking in which no transition is enabled. The behavior of a Petri net is defined as the set of all its possible runs.

\medskip

WF-nets are a special subclass of Petri nets designed for modeling workflow processes. 
Each WF-net has two special places: $i$ and $f$. These places are used to mark the beginning and the ending of executing one case of a workflow process.

Formally, a marked Petri net $N=(P,T,F, m_0)$ is called a  \emph{workflow net  (WF-net)} iff
\begin{itemize}
	\item
	there is one source place  $i\in P$ and one sink place $f\in P$ s.~t.
	$\pre{i}=\post{f}=\emptyset$;
	\item
	every node from $P\cup T$ is on a path from  $i$ to $f$;
	\item
	the initial marking $m_0$ in $N$ contains the only token in its source place $i$.
\end{itemize}

\begin{figure}[ht]
	\centering
	\includegraphics[width=\textwidth]{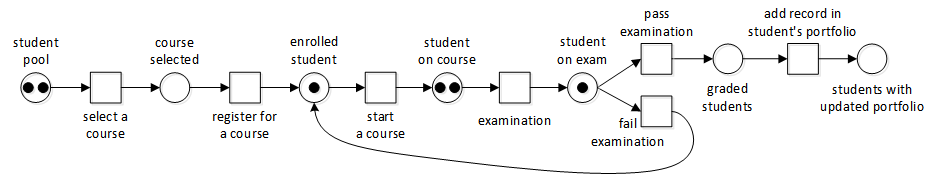}
	\caption{Workflow net modeling the process of  a MOOC study}
	\label{FIG:1}
\end{figure}

Fig.~\ref{FIG:1} provides an example of a WF-net model, where a case is the process of one student learning a specific MOOC course.  
Here, after selecting the course, a student registers for it, starts the course, and takes the exam. Then there is a fork. If the student fails the exam, he/she repeats the course; if the student successfully passes the exam, he/she receives corresponding record in his/her portfolio. In this simple workflow net, firing of a transition transfers a token from its current place, which represents its local state, to a new one.

The marking shown in this figure represents some intermediate state, when there are two students in the student pool (the source place), one student is to start the course, two students are studying the course, and one is taking the exam.

Note that in this workflow model, students are represented by black dot tokens and cannot be distinguished. Thus, the model demonstrates how many students are currently taking the course and at what stage they are now. For example, it is possible to analyze some performance issues using such a model. This may be not enough. So, we extend the capabilities of workflow modeling to express more perspectives and details of the learning process.

Next, we show how CPNs, in which tokens carry data values from different domains, can be used to model the MOOC. 

\subsection{Modeling Learnflow with Colored Petri Nets}

In our case, the further extension of the classical Petri net model is CPN, where different colors of tokens are different students. Fig. 2 presents an example of a CPN modeling students' trajectories in MOOC. This example is made using CPN Tools\footnote{http://cpntools.org/}. It models the study on one particular course, \ie how the student passes through all stages of the  course. Here tokens representing students are  tuples of the form $(id, r)$, where $id$ is a student identifier (or ID), and $r$ is a record in a student's portfolio, containing learning trajectory data such as completed courses and grades. 
Thus, unlike the classical Petri net model, in CPN model we have additional information about the students. This information is presented in the data structures defined in the declarative part of the model on the left in Fig. 2.  
Declarative parts in CPNs define variables, constants, and functions that can be used to describe operations associated with transitions firings. 
For example, the transition, `pass exam',  adds completed course data to a student's portfolio.
CPNs also provide boolean conditions (guards), extending the definition of the transition enabling. For example, the transition, `register for a course', fires only if the course `23' (may be a prerequisite for the selected course) is already completed.

The net component in the CPN is a Petri net with places, transitions, and arcs equipped with additional labels (shown on the right in Fig.~2). The marking of the CPN maps its places to multisets of colored tokens, and its transitions can fire in different modes, depending on the tokens involved in the firing.  In our example in Fig.~2, in the current marking there are two students with IDs `1' and `34' in the `student pool', where the first one has courses `1' and `2' in his/her portfolio, and the portfolio of the second student is empty.   There are also one student enrolled, two students on the course, and one student taking the exam.  In this marking, the transition `select course' can fire in two modes ---  either for the student with ID `1', or for the student with ID `34'. This is implemented by binding the variable $id$ with the value  `1' or `34' correspondingly.

\begin{figure}[ht]
	\centering
	\includegraphics[width=\textwidth]{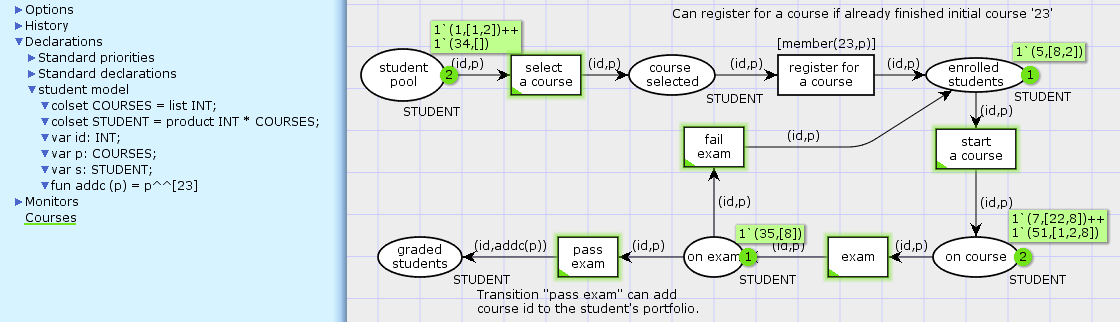}
	\caption{Colored Petri net modeling the process of  a MOOC study}
	\label{FIG:2}
\end{figure}

Now let us define CPNs more precisely. The following 
definition is based on the classical one~\citep{JensenBook}.

In CPNs tokens are attached with typed data values (colors). Each place in a CPN is equipped with a type indicating the type of tokens it can carry. Thus, a type (a color set) is a set of values. Let $\Type$ be a set of types, ${\cal U} =\bigcup_{S\in \Type} S$ --- the universe of typed data values.   Let then $\Con$ be a set of (typed) constants with values in  ${\cal U}$, $\Var$ --- a set of (typed) variables, and
$\Expr$ --- a language  for arcs expressions over  $\Var$
and $\Con$, s.t. for any type-consistent variable binding
(variable evaluation) $b : \Var \to {\cal U}$   the value $e(b) \in \cal U$ of an expression $e \in \Expr$ is defined. Note that for a concrete model constants, variables and operations used in expressions are defined in the declarative part of the CPN.

By abuse of notation for an expression $e\in \Expr$ by $\Type(e)$ we denote the type of $e$, i.e. the type of its values.
There is also a language $\Guard$ of  transition guards over  $\Var$ and $\Con$ and (without loss of generality) we suppose that for any binding $b: \Var \to \cal U$   the boolean value $g(b) \in \{\true,\false\}$ of the guard expression
$g \in \Guard$ is defined.

Then a CPN is a tuple ${\cal N} = (P,T,F,\tau,\gamma, \epsilon)$, where
\begin{itemize}
	\item $(P,T,F)$ is a Petri net;
	\item $\tau: P \to \Type$ assigns types to places;
	\item $\gamma: T \to \Guard$ is a partial function assigning guards to transitions (if there is no guard, it is supposed to be $\true$);
	\item $\epsilon: F \to \Expr$ is an {\em arc labeling} function; the type of $e$, labeling an arc $(p,t)$ or $(t,p)$, should coincide with the type of the place $p$ incident to $e$.

\end{itemize}

A marking $m$ in CPN is a function that maps each place $p \in P$ to a multiset of tokens $m(p) \in {\cal M}(\tau(p))$. 
For a CPN, we distinguish an initial marking $m_0$. A marked CPN is a CPN with its initial marking. 

Let ${\cal N} = (P,T,F,\tau,\gamma, \epsilon)$ be a CPN. A binding $b: \Var \to {\cal U}$ for a transition $t\in T$ is a partial function that assigns a type-consistent value $b(v)$ to each variable $v\in \Var$ which occurs in the guard $\gamma(t)$, or in  expressions of  arcs incident to $t$. 

A binding determines a mode of transition firing. 
An evaluation $\epsilon(p, t)(b)$ determines token demands (multiset of tokens) on $p$ for $t$ to be enabled with the binding $b$, and the multiset of tokens that the transition $t$ removes from the place $p$ when $t$ occurs with the binding $b$. $\epsilon(t, p)(b)$ determines the multiset of tokens added to an output place $p$. More precisely, a transition $t$ is enabled in a marking $m$ w.r.t a binding $b$ iff for all $p \in P$, $\epsilon(p, t)(b) \subseteq m(p)$. An enabled transition $t$ fires in a marking $m$ yielding a new marking $m'$, such that for all places $p$, $m'(p) =(m(p) - \epsilon(p, t)(b)) + \epsilon(t, p)(b)$.

Thus, the CPN model allows us to represent the control-flow of students' activities when studying a course, depending on their data (grades, passed courses, etc.). Furthermore, student data is localized, i.e. available only for the student in his/her current state. In order to model several concurrent processes performed by  the same student, for example,  when studying several courses, we need to further expand our modeling capabilities.

\subsection{Modeling Learnflow Using Petri Nets with Reference Tokens}

When a student studies only one course, his/her data can be 'attached' to him/her (localized). But when a student is studying a multi-course program, it is important that his/her current records are available to all components of the supporting information system. To model this, we use tokens that are not data values, but are references to shared data.

The idea of extending Petri nets with pointer tokens goes back to R.~Valk~\citep{Valk98} who considered tokens referring to elementary Petri nets, representing individual agents. Developing this concept of 'net-within-net' in terms of reference semantics M.~K{\"o}hler~\citep{Kohler2001} used Java inscriptions to create nets with binding to reference variables. On the other hand, \cite{MontaliRivkin} considered DB-nets with tokens referring to a database. In \cite{CarLomRiv2020}, Petri nets with tokens referring to global data were used for modeling trading systems.

In this paper, we present PNRD --- the extension of the Petri net model with tokens allowed to be references and very clear and intrinsic semantics. 

Fig. 3 presents a PNRD example, where tokens are referenced to objects in a database, providing the manager of the course with data from the student portfolios.

\begin{figure}[ht]
	\centering
	\includegraphics[width=0.85\textwidth]{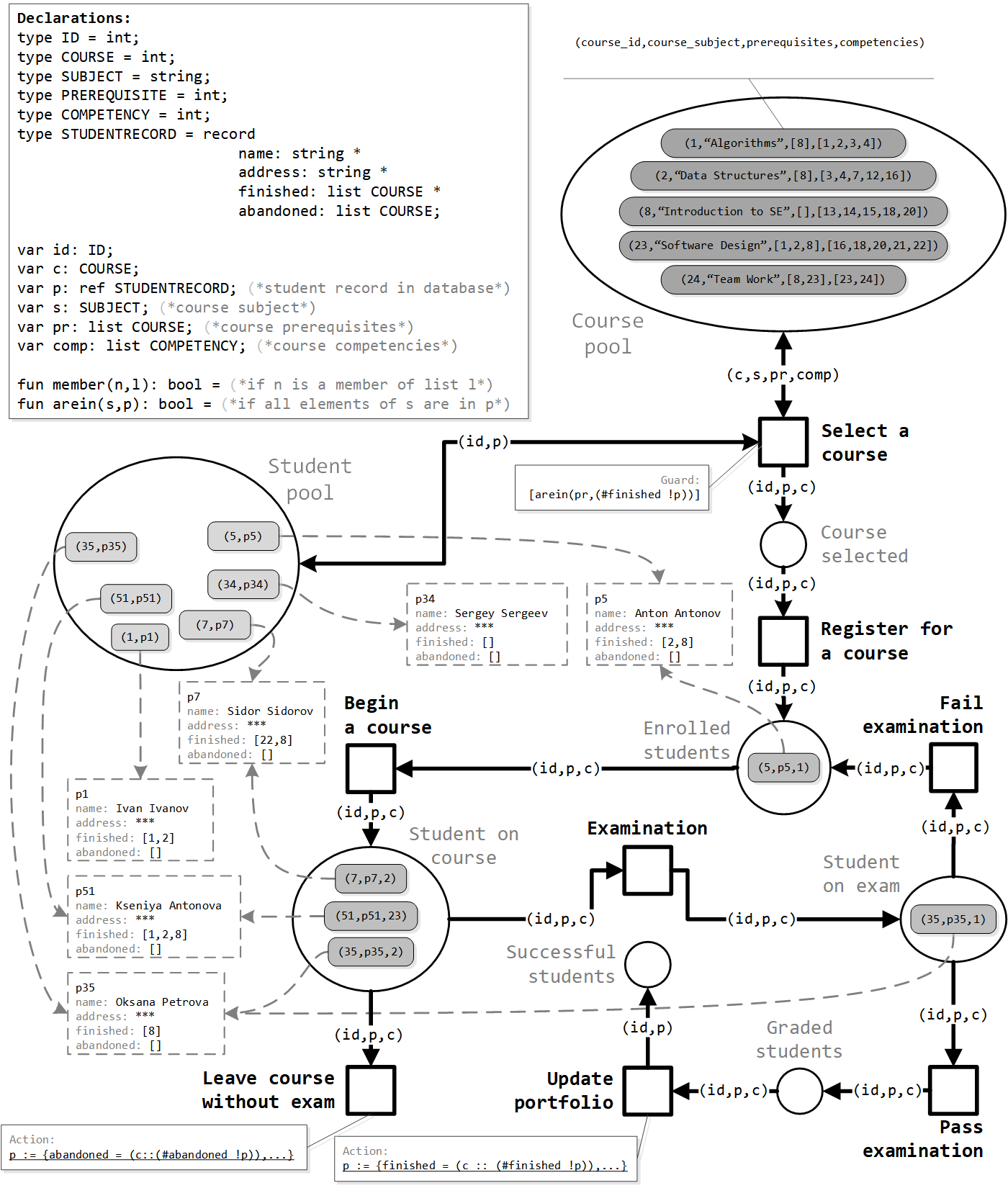}
	\caption{Example of PNRD for modeling MOOC elements of student MOOC study}
	\label{FIG:3}
\end{figure}

The model describes the education process, when many students take many courses. The pool of students contains students represented by their ID and a reference to their portfolio. Thus to identify students, both colors and reference to the portfolio are used to model learning trajectories. When the student studies various courses in parallel, there are several tokens related to the same student, where each token is presented by a tuple of the student ID, reference to his/her portfolio,  and the course ID he/she is studying. Hence, it is crucial to use references mechanism to model the student learning trajectory across multiple courses. 

The pool of courses contains tokens for the available courses. 
Each token in this pool is a tuple consisting of  course ID, course name, and prerequisites --- list of courses to learn first. 
 Accordingly, a student can choose  a course if the set of prerequisites for the course  is included in the set of completed courses in his/her portfolio. 

When a student enrolls in the course, the record of successful enrollment is added to the portfolio. After firing the transition, 'start a course', a student is placed 'on the course' and begins training.  The information about this course (course ID) is added to his/her portfolio. 

For a token residing in  the place 'student on course', there are two possibilities. If  the transition 'examination' fires, the token is carried over to the place 'student on exam' representing students taking exams. After that appropriate data on the exam and the final grade is added to the portfolio, or a student may fail the exam and the corresponding token  is moved to the place 'enrolled student' to start the course again and try to complete it successfully. Another possibility for the token in the place 'student on course' is to enable  the transition 'leave course without exam', which removes the token. This means that a student has dropped out of the course study without taking the exam and this information is added to his/her portfolio. 

In this model, the successfully completed course information is added to the portfolio after the student has passed the exam. In fact,  information about the student's state can be added at all stages of the course study,  for example, data on enrollment in the course, the results of intermediate tests , etc. These details are omitted to avoid reloading the model. This example shows how an online course trajectory can be modeled in the context of using references to objects that contain data. Also note that the modeling technique used in this example allows us to model \emph{adaptive learning} where course prerequisites can be based on intermediate results for another course, not just completed course records.

\medskip 
Now we define PNRD more formally. First, PNRD is a CPN extension.

The main feature here is that tokens in PNRD can refer to  \emph{global data}. Without loss of generality, we suppose that values of local and global data are elements of the universe $\cal U$. So we extend the set of types $\Type$ with pointer types in such a way that for  type $\tau \in \Type$, $\tau\!\!\uparrow$ will denote the type of pointers to $\tau$-values. The set of pointer types will be denoted by $\Type\!\!\uparrow$.

By $\Pnt$ we denote a set of reference constants, or pointers --- names of pointers to global values from ${\cal U}$. Note that  global data is only accessible through pointers. Of course, the values pointed to  are subject to change. The current evaluation of pointers that defines the state of global data is  the function $s : \Pnt \to {\cal U}$ that maps pointers to global values. By $\cal S$ we will denote the set of all global states.

For simplicity of presentation, we suppose that tokens in PNRD are tuples of values from  ${\cal U}$ and pointers. In most cases, this is sufficient, e.g. in our example of modeling MOOC study in Fig.~\ref{FIG:3}, a token is a pair of a student ID and a pointer to his/her portfolio, or a triple of a student ID, a pointer to his/her portfolio, and a set of course IDs.
Note that allowing tokens that carry data with a more complex structure is not a problem, but the presentation would be more cumbersome.
As in CPN, each place in PNRD is equipped with a type indicating the type of tokens it can carry. 

Similar to CPN, in arc expressions we use typed variables. But now additionally to variables from $\Var$ (with values in $\cal U$) there are  reference variables whose values are pointers from $\Pnt$. The set of reference variables will be denoted by $\Var\!\!\uparrow$. 

Thus, arc expressions in PNRD are built over constants from $\Con$, pointers from $\Pnt$,  and variables from $(\Var \cup \Var\!\!\uparrow)$. Of course, the type of $e$, labeling an arc $(p,t)$ or $(t,p)$, should coincide with the type of  place $p$ incident to $e$. It means that an arc expression  in PNRD is a tuple of expressions $e=(e_1, \dots, e_k)$, where $\tau(e_i)\in (\Type \cup \Type\!\!\uparrow)$ coincides with the type of the $i$-th component of tokens in the corresponding place. 
Overriding the notations for PNRD, by $\Expr$ we  denote  a set of arc expressions, and by 
$\Guard$ --- a set of  transition guards (Boolean expressions) over  $\Var$, $\Var\!\!\uparrow$,  $\Con$, and $\Pnt$. Recall that for a concrete model, its constants, pointers, variables and operations used in expressions are defined in the declarative part of the CPN. Naturally, we do not allow operations on pointers, so an expression with a type from $\Type\!\!\uparrow$(reference expression) is just a pointer, or a variable from $\Var\!\!\uparrow$.

The value of an arc  or guard expression   depends on the evaluation of the variables  (binding) and the  state of the global data. The value of a  reference expression is completely determined by the binding that maps the variables from $\Var\!\!\uparrow$ to pointers. As for value expressions, given a binding $b: (\Var \cup \Var\!\!\uparrow) \to ({\cal U}\cup \Pnt)$ and a global state $s \in \cal S$, the guard or expression  value is computed by substituting  value $b(v)$ for each variable $v\in \Var$, and  value $s(b(w))$ for each variable $w\in \Var\!\!\uparrow$. The  value of a tuple expression is then the tuple of the values of its components. Further we write $\nu(e,b,s)$ for the value of expression $e$ over binding $b$ and global state $s$.

After we have overridden the types, expressions and guards for PNRD, the definition of PNRD is almost the same as for CPN. The difference is that transitions in PNRD can perform operations on global data, such as  adding, deleting, or modifying records in a database. So, we suppose there is also a domain-specific language $\Op$ over  $\Var$, $\Var\!\!\uparrow$,  $\Con$, and $\Pnt$ for global data operators, such that for each operator expression $e\in \Op$ its semantics in the form of state transformer $o(e): {\cal S} \to {\cal S}$ is defined.

\medskip

In the context of all the above, PNRD  is a tuple ${\cal N} = (P,T,F,\tau,\gamma, \theta, \epsilon)$, where
\begin{itemize}
	\item $(P,T,F)$ is a Petri net;
	\item $\tau: P \to (\Type \cup \Type\!\!\uparrow)^*$  assigns tuples of types from $(\Type \cup \Type\!\!\uparrow)$ to places;
	\item $\gamma: T \to \Guard$ is a partial function assigning guards to transitions ($\true$ by default);
	\item $\theta: T \to \Op$ is a partial function assigning operators to transitions ($\skp$ by default);
	\item $\epsilon: F \to \Expr$ is an {\em arc labeling} function; the type of $e$, labeling an arc $(p,t)$ or $(t,p)$, should coincide with the type of the place $p$ incident to $e$.

\end{itemize}

A marking $m$ in PNRD is a function that maps each place $p \in P$ to a multiset of tokens $m(p)$ of the corresponding type $\tau(p)$. In contrast to CPN,  in PNRD model a state is defined not just as a marking, but as a marking along with the global data state, which is determined by 
the evaluation of pointers $s : \Pnt \to {\cal U}$ which maps pointers to global values. Then a state in PNRD is a pair $\sigma =(m,s)$, where $m$ is a marking, and $s$ is a global data state.

Now we come to defining the firing rule for PNRD.

Let ${\cal N} = (P,T,F,\tau,\gamma, \theta, \epsilon)$ be a PNRD, $\sigma= (m,s)$ --- its current state. A binding $b: (\Var \cup \Var\!\!\uparrow) \to ({\cal U}\cup \Pnt)$ for a transition $t\in T$ is a partial function that assigns the (type consistent) value $b(v) \in {\cal U}$ to each variable $v\in \Var$ and, accordingly, the pointer $b(w)\in \Pnt$ to each variable $w\in \Var\!\!\uparrow$ that appear in  guard $\gamma(t)$, or in  expressions on  arcs incident to $t$. 
As in CPN, a binding determines a mode of transition firing. 

An evaluation $\nu(\epsilon(p, t),b,s)$ determines token demands (multiset of tokens) on $p$ for $t$ to be enabled with the binding $b$, and the multiset of tokens that the transition $t$ removes from the place $p$ when $t$ occurs with the binding $b$. It also changes global state $s$ to a new global state according to the operator $\theta(t)$. An evaluation $\nu(\epsilon(t, p),b,s)$ determines the multiset of tokens added to an output place $p$. More precisely, a transition $t$ is enabled in a state $(m,s)$ w.r.t a binding $b$ iff for all $p \in P$, $\nu(\epsilon(p, t),b,s) \subseteq m(p)$. An enabled transition $t$ fires in a state $(m,s)$ yielding a new state $(m',s')$, such that for all places $p$, $m'(p) =(m(p) - \nu(\epsilon(p, t),b,s)) + \nu(\epsilon(t, p),b,s)$, and $s' = o(\theta(t))(s)$.

Similar to classical Petri nets, a PNRD run is a sequence of firings $\sigma_0\to \sigma_1\to \dots$ starting from the initial state $\sigma_0$. 
 
The PNRD behavior is specified as a set of all its possible runs.

\subsection{Modeling Teamwork using PNRDs}

MOOCs can be individual pursuits, but their real effectiveness as a learning environment comes from the collaboration and teamwork that they make possible. 
We believe that cross-person interactions are at the center of the education process. 
It is very important for the lecture to monitor and analyze the interaction of students in order to detect and solve problems that arise. Leranflow models can help with this, but it is impossible to trace these aspects by simulating learning just for individual students, and excluding  teamwork and organization of  teamwork on a project.

We will now show that PNRD models allow us to express student interactions. Fig. 4 shows an example of  student collaboration work, namely the implementation of a group project while studying an online course. 

\begin{figure}[ht]
	\centering
	\includegraphics[width=0.95\textwidth]{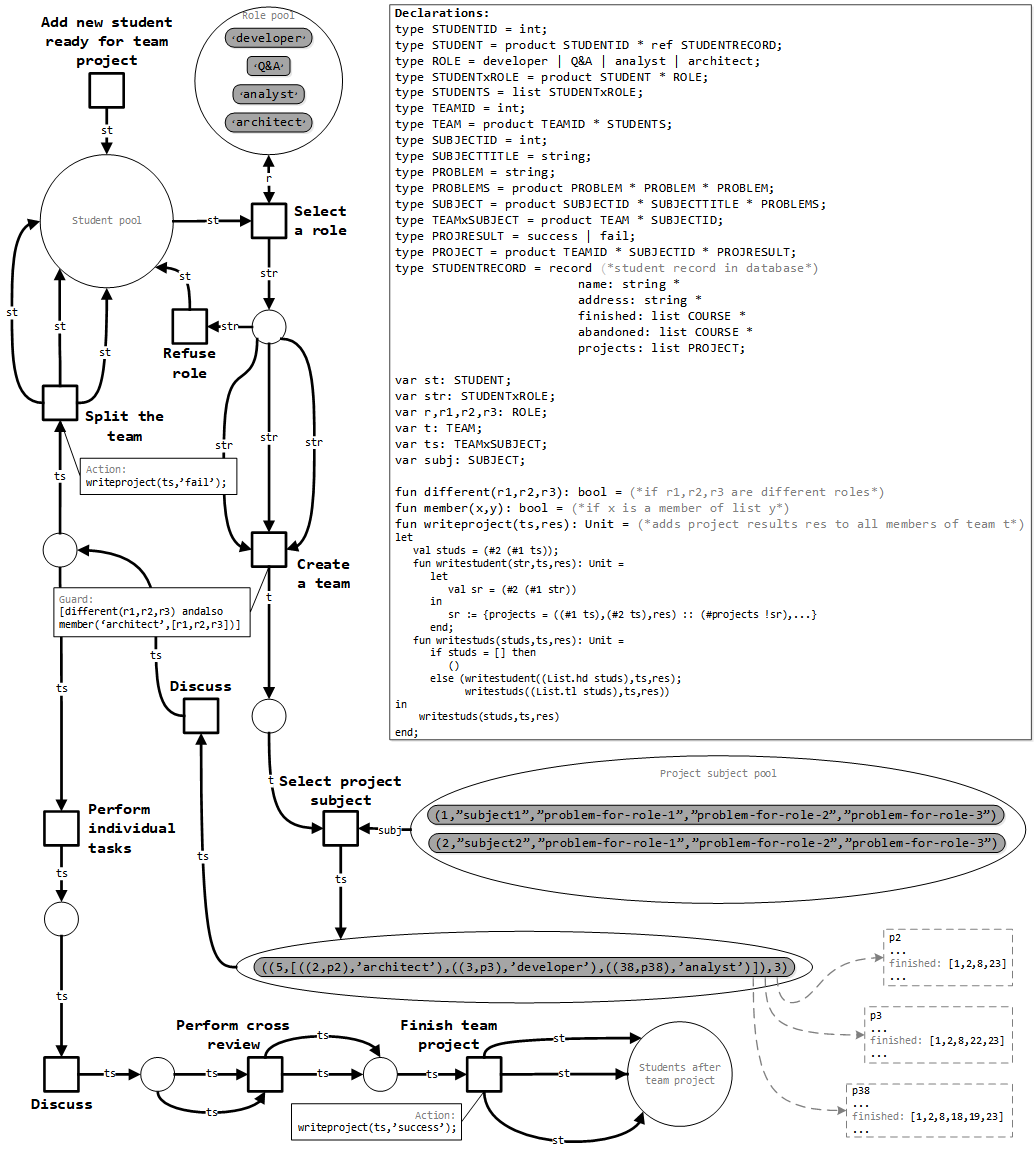}
	\caption{PNRD representing teamwork workflow. A list of items in square brackets indicates the set of these items}
	\label{FIG:4}
\end{figure}

A student in the place `Student pool' is encoded with a tuple (id, p), where `id' is the student ID and `p' is a reference to  the student's portfolio. After firing of the transition `select role' , a student can select one of  the possible roles in the project from the place `role pool'. Accordingly, at this stage, a student is identified as a tuple (id, p, r) ---  the student's ID, reference to the portfolio, and the selected role `r'. 

Thus, tokens in the place `student with selected role' represent  students with selected roles ready to join a team. Then a team is created from these students, labeled with the team ID `tid'. To create a team, all the roles of its members must be different. If the transition `refuse role' fires, a token returns to `student pool'. 
The `select project' transition firing simulates a team's selection of a project theme from the pool of project subjects and adds the subject identifier `sid' to the tuple representing the team.
Each project can be selected by only  one team; if the team stops working on the project,  `sid' of the project is returned to the project's subject pool. 

Then, students discuss the work on the project.
When the transition `Discuss' has fired,   the team token changes to the state where the choice is possible. The students may decide to cancel the project and return to `Student pool' by splitting their team. Another option is to complete individual assignments and then discuss the results of the project. After that, in the place `project discussion'   tokens appear  for teams that  cross-review in pairs and then complete the project. As a result, a record of a successful team project is added to the portfolio. 
										
\section{Discussion}\label{discussion}
\subsection{Key Findings}

Modern education is supported by different types of information systems.
Both online and conventional education models assume now that students access course data and submit their solutions in some type of a computerized system.
The spread of such computer-aided education practices gives new opportunities to monitor, analyze, and improve education.
This paper investigates how different aspects of the education process can be formally modeled using a mathematical language of Petri nets.

In the paper, we emphasize that \textbf{education is a distributed process} in which multiple agents (students and professors) can perform interdependent actions, as well as communicate and exchange data both synchronously and asynchronously.
Students attend courses, submit individual task solutions, perform team projects.
All these complex activities are supported by information systems, and thus can be analyzed using mathematical methods.
To do this, the first step is to model.

In this paper, \textbf{we present a Petri net extension} to model the education process in a MOOC.

First, we identified the important aspects of education process which should be modeled.
Typically, each student is allowed to attend more than one course at the same time.
Student performance on each of the courses can influence his or her results on other courses.
Moreover, subjects learned in one course can be credited in other courses.
Teamwork is of key importance in modern education practices as well.
Students are linked to individual grade records etc.
Classical Petri nets can not be used to model these aspects.

So (second), we proposed a new formalism supporting all these phenomena.
An education process model in this new formalism can be used with different goals.
An expert may use it \textbf{to understand the real process} better.
It is possible \textbf{to identify difficulties} that a student may face. 
For example, if combined with additional student data, typical student trajectories can be found and projected onto a model together with outliers.
Consequently, the model enables us to determine at what stages of the course or what types of work the student has a problem with and define the reasons for these problems.
On the other hand, this could be a course management problem due to the complex and distributive nature of teamwork in MOOCs.

The modeling of a massive learnflow with respect to the students' teamwork and an individual student trajectory enables \emph{better identification of bottlenecks due to communication}. 
Of high importance is to eliminate these early in the process.
Students often interconnect with teammates during a course study.
It is essential to combine individual student motivation, their ability to work independently and effectively with benefits of students' online \textbf{collaborative work}.
Individual aspects of MOOC learners should be taken into account when individuals form a team in project-based education process.
Besides supporting the faster achievement of main technical goals, this may help in collaborative outcomes.

The analysis of a student's learning trajectory can be based simply on modeling a single workflow.
However, a model that shows how the student passes from class to class within a single discipline until the final exam does not reflect the full education process in computerized MOOC or LMS.
In reality, a student takes more than one course at once.
Studying online, a student goes through many courses, which may have common subjects.
Consider, for example, how software testing can be taught in programming, software engineering, and other courses.
So, do our students really need to learn a subject twice or their results can be just taken from one course to another?
Prerequisites are also important to model.
A student plans to study in course ``A'', however, to enroll, the student needs to have the skills from course ``B''.
By using our proposed \textbf{model with references to data}, which courses a student has taken can be identified and the records are in the student's portfolio. 
Thus, we can model inter-course causal dependencies and use a single subject in many different courses.
Our new formalism allows for performing \textbf{direct process verification} with respect to these communication aspects.

References to data in the proposed model can be applied for a nonlinear \textbf{adaptive course design}.
A student may receive subject credits from different disciplines simultaneously.
The individual \textbf{learning trajectory} can be many more \textbf{of declarative nature} than prescriptive.
An expert has to specify what can not be done (using course models and prerequisites) rather than what to do.
Using references to data related to process objects, experts are able to develop a simpler and less overloaded model.
Again, process models can be tested using simulation or even verified, and only then implemented in a MOOC or LMS.

\subsection{Conclusions and Future Work}

In this paper, we investigated the development of a multi-agent model of the e-learning process. 
We proposed a new mathematical formalism based on Petri nets which allows for modeling advanced aspects of the learning process: inter-agent communication, complex causal dependencies, distributed behavior of agents.
One can model teamwork using our formalism as well.

This article is one of the first studies using Petri-net-based models that take teamwork into consideration. 
The model of team workflow may be a prototype for designing an e-learning environment for the analysis of MOOC performance and to organize an efficient environment for all participants (learners and teachers) of MOOCs and LMSs. 
The models based on the proposed formalism provide a holistic view, capturing various aspects of the online education process at an appropriate level of abstraction.

Our MOOC learning model enables a visual representation of the e-learning process. 
Note that visual analytics is used wide in business process management, software and technical process analysis \citep{Keim0028506,KeimMSZ18,kotylev2020}.
It is possible to determine where bottlenecks occur on different levels within detailed information about courses. 
Course management is able to monitor education processes and identify bottlenecks in education and organizational processes for individual students, whole courses, or even multi-course learning environments.

It is important to highlight problems that will be a subject for \textbf{further research}.

The essential part of MOOC research is in the field of software and technical support. 
For modern complex e-learning information systems, it is necessary to rationally optimize the available resource usage to ensure that MOOCs work efficiently, without communication difficulties between the participants. 
The proposed formalism can support the technical organization of such systems. 
However, the modeling formalism is general.
Thus, when dealing with each concrete system models need to be adapted to the features of a specific MOOC platform. 
One of planned future projects is to perform case studies based on particular MOOC systems.

Other possible direction of future work is to enrich learning process models with learning system data.
In particular, all these systems record event logs with technical information and user activity event.
Fortunately, process mining \citep{aalst16book} flourished in last years on rich plains of data.
Using methods of this field it is possible to align models made by experts with the real process data \citep{CarmonaDSW18book}.
Moreover, process monitoring \citep{MaggiFDG14} facilities are considered very useful and promising.
Process mining was implemented for the analysis of various problems in education, as there is a large amount of LMS data \citep{Bogarin2018}.
Thus, we believe the combination of our formalism with process mining will give us new opportunities for visualizing, monitoring, compliance and conformance checking in e-learning environments.

\printcredits

\end{document}